\documentclass[superscriptaddress,twocolumn]{revtex4-1}
\usepackage{amsmath,sidecap,epsfig}
\usepackage{hyperref}
\usepackage[brazil]{babel}
\usepackage[latin1]{inputenc}
\usepackage{graphicx}

\begin{document}

\title{The pasta phase and its consequences on neutrino opacities}

\author{M. D. Alloy}
\email[E-mail me at:]{alloy@uffs.edu.br}
\affiliation{Universidade Federal da Fronteira Sul, Chapecó, SC, CEP 89.812-000, Brazil.}

\author{D. P. Menezes}
\email[E-mail me at:]{debora@fsc.ufsc.br}
\affiliation{Depto de Física CFM, Universidade Federal de Santa Catarina, 
Florianópolis, SC CP.476, CEP 88.040-900, Brazil}

\begin{abstract}
In this paper, we calculate the diffusion coefficients that are related to the
neutrino opacities considering the formation of nuclear pasta 
and homogeneous matter at low densities. Our results show that the mean free 
paths are significantly altered by the presence of nuclear pasta in stellar 
matter when 
compared with the results obtained with homogeneous matter.
These differences in neutrino opacities certainly influence the
Kelvin-Helmholtz phase of protoneutron stars and consequently the results
of supernova explosion simulations.
\end{abstract}

\keywords{protoneutron stars, neutrino opacities}
\date{\today}
\maketitle
\twocolumngrid

\section{Introduction}

When massive stars ($8~M_\odot < M < 30~M_\odot$ ) exhausts its fuel supply, 
the forces that support the stars core quickly retreat, and the core is almost 
instantly crushed by gravity, which triggers a type II supernova
explosion. The remnant of the gravitational collapse of the core
of a massive star is a compact star or a black hole, depending on the 
initial condition of the collapse. Newly-born protoneutron stars (PNS) are hot 
and rich in leptons, mostly $e^-$ and $\nu_e$  and have masses of the order of
$1-2~M_\odot$ \cite{Lattimer2004,keil1995}. During the very beginning of the 
evolution, most of the binding energy, of the order of $10^{53}$ ergs is 
radiated away by the neutrinos. 

The composition of protoneutron and neutron stars remains a source of 
intense speculation in the literature. Whether their internal structure is
formed by nucleons and leptons, by other light baryons and leptons, 
by baryons, leptons and quarks (bearing or not a mixed phase), by baryons, 
leptons and kaons or by other possible composition, is still unknown. 
The neutrino-signals detected by astronomers
can be used as a constraint to infer protoneutron star composition 
\cite{keil1995,pons2001}. For the same purpose, theoretical
studies involving different possible equations of state obtained for all sorts
of matter composition have to be done because the temporal evolution of the 
PNS in the so-called Kelvin-Helmholtz epoch, during which the remnant compact 
object changes from a hot and lepton-rich 
PNS to a cold and deleptonized neutron star depends on two key ingredients:
the equation of state (EoS) and its associated neutrino opacity at supranuclear
densities \cite{pons1999,pons2001}. 

Neutrinos already present or generated in the PNS hot matter escape by
diffusion (not free streaming) because of the very high densities and
temperatures involved. The neutrino opacity is calculated from the scattering 
and absorption reactions that take place in the medium and hence, related to 
its mean free path, which is of the order of 10 cm and much smaller than the 
protoneutron star radius \cite{burrows1986}. In the diffusion approximation
used to obtain the temporal evolution of the PNS in the Kelvin-Helmholtz phase,
the total neutrino mean free path depends on the calculation of diffusion 
coefficients, which, in turn, depend on the chosen EoS.
At zero temperature no trapped neutrinos are left in the star because their 
mean free path would be larger than the compact star radius.

A complete equation of state capable of describing matter ranging from very 
low densities to few times saturation density and from zero temperature to 
around 50 MeV is a necessary step towards the understanding of PNS
evolution. The constitution of the PNS crust plays a definite role in the 
emission of neutrinos. For this reason, the pasta phase, present in very
low nuclear matter as the crust of PNS are included in
the investigation of the neutrino opacity in the present work.

A few words on the pasta phase follow. It is a frustrated system 
\cite{pethick,horowitz2004,horo,maruyama,watanabe05} present at densities 
of the order 
of 0.006 - 0.1 fm$^{-3}$ \cite{pasta1} in neutral nuclear matter and 
0.029 - 0.065 fm$^{-3}$ 
\cite{bao,pasta2} in $\beta$-equilibrium stellar matter, where a competition 
between the strong and the electromagnetic interactions takes place. 
The basic shapes of these complex structures were named \cite{pethick} 
after well known types of cheese and pasta: droplets = meat balls (bubbles = 
Swiss cheese), rods = spaghetti (tubes = penne) and slabs = lasagna, for 
three, two and one dimensions respectively. 
The pasta phase is the ground state configuration if its free energy per 
particle is lower than the corresponding to the homogeneous phase at the same 
density. 

The evolution of PNS and simulation of supernova explosion have already been 
considered for different matter compositions, some with the inclusion of the 
pasta phase \cite{pons1999,pons2001,horowitz2004,sonoda2007,watanabe09}.
From \cite{pons1999,pons2001} one can see that the transport properties are 
significantly affected by the presence or absence of hyperons and of the mixed
phase in hybrid stars. In \cite{horowitz2004} the linear response of the
nuclear pasta to neutrinos was calculated with a semi-classical simulation and 
the muon and taon neutrinos mean-free path were described by the static
structure factor of the pasta evaluated with Metropolis Monte Carlo simulations.
In \cite{sonoda2007} rod-like (two dimensions) and slab-like (one dimension) 
pasta structures were included in the calculation of neutrino opacity within
quantum molecular dynamics. A very interesting conclusion was that the pasta
phase occupies 10-20\% of the mass of the supernova core in the later stage of
the collapse.

In the present work we investigate the influence of the pasta phase on the
neutrino opacity by showing the effects on the diffusion coefficients.
The pasta phase is calculated with the coexistence phases method (CP) in a mean
field approximation \cite{pasta1,pasta2,pasta_alpha}. We consider only 
nucleons and leptons in the EoS in $\beta$-equilibrium. In the pasta structure 
only electron neutrinos are considered. 

In section II we present the formalism used to obtain the equation of state,
in section III the recipe used for the construction of the pasta phase is 
outlined, in section IV the expressions used to calculate the neutrino cross
sections and related mean free path are given and in section V our results are
shown and the main conclusions are discussed.

\section{Formalism}

We consider a system of protons and neutrons with mass $M$ interacting
with and through an isoscalar-scalar field $\phi$ with mass $m_s$, an
isoscalar-vector field $V^\mu$ with mass $m_v$ and an isovector-vector
field ${\bf b}_\mu$ with mass $m_\rho$ described by the well known non-linear
Walecka model (NLWM) \cite{sw}. We impose $\beta$-equilibrium and charge
neutrality with neutrino trapping at finite temperature. At zero temperature no
neutrinos are left in the system.

The Lagrangian density reads
\begin{equation}
\mathcal{L}=\sum_{j=p,n}\mathcal{L}_j+\mathcal{L}_\sigma+
\mathcal{L}_\omega+\mathcal{L}_\rho+\sum_{l=e,\nu}\mathcal{L}_l,
\end{equation}
where the nucleon Lagrangian reads
\begin{equation}
\mathcal{L}_j=\overline{\psi}_j\left[\gamma_\mu iD^\mu-M^{*}\right]\psi_j,
\end{equation}
were $M^{*}=M-g_s\phi$ is the effective baryon mass and
\begin{equation}
iD^\mu=i\partial^\mu-g_vV^\mu-\frac{g_\rho}{2}{\bf\tau\cdot b^\mu}.	
\end{equation}

The meson Lagrangian densities are given by
\begin{eqnarray}
\mathcal{L}_\sigma&=&\frac{1}{2}\left(\partial_\mu\phi\partial^\mu\phi-m_s^2\phi^2-\frac{1}{3}\kappa\phi^3-\frac{1}{12}\lambda\phi^4\right),\\
\mathcal{L}_\omega&=&\frac{1}{2}\left(-\frac{1}{2}\Omega_{\mu\nu}\Omega^{\mu\nu}+m_v^2V_\mu V^\mu \right),\\
\mathcal{L}_\rho&=&\frac{1}{2}\left(-\frac{1}{2}{\bf B_{\mu\nu}\cdot B^{\mu\nu}}+m_\rho^2{\bf b_\mu\cdot b^\mu}\right),
\end{eqnarray}
where $\Omega_{\mu\nu}=\partial_\mu V_\nu-\partial_\nu V_\mu$ and
${\bf B_{\mu\nu}}=\partial_\mu {\bf b_\nu}-\partial_\nu {\bf b_\mu}-g_\rho({\bf b_\mu}\times {\bf b_\nu})$.  
The lepton Lagrangian densities read
\begin{equation}
\mathcal{L}_l=\overline{\psi}_l\left[\gamma_\mu i\partial^\mu-m_l\right]\psi_l,
\end{equation}
where $m_e$ is the electron mass and the neutrino mass is $m_\nu=0$.

The parameters of the model are three coupling constants $g_s$, $g_v$ and 
$g_\rho$ of the mesons to the nucleons,
the nucleon mass $M$, the electron mass $m_e$, the masses of the mesons $m_s$, 
$m_v$ and $m_\rho$ and self-interacting coupling constants $\kappa$
and $\lambda$. The numerical values of the parameters used in this work and 
usually referred to as NL3 \cite{nl3} are shown in table 
\ref{tab:parameters_set}. They are fixed in such a way that the main nuclear 
matter bulk properties are the binding energy equal to 16.3 MeV at the 
saturation density 0.148 fm$^{-3}$, the compressibility is 272 MeV and the 
effective mass at the saturation density is 0.6 M. 

\begin{widetext}
\begin{table}[hb]
\begin{tabular}{llllllllllll}
\hline
Model & $g_s$ & $g_v$ & $g_\rho$ & $M$ & $m_e$ & $m_s$ & $m_v$ & $m_\rho$ & $\kappa/M$ & $\lambda$ \\
\hline
\hline
NL3   & 10.217 & 12.868 & 8.948 & 939.0 & 0.511 & 508.194 & 782.501 & 763.0 & 4.377 & -173.31 \\
\hline
\hline
\end{tabular}
\caption{\label{tab:parameters_set} Parameters set used in this work. All masses are given in MeV.}
\end{table}
\end{widetext}

From the Euler-Lagrange formalism we obtain the equations of motion for the 
nucleons and for the meson fields:
\begin{eqnarray}
&\nabla^2&\phi=m_s^2\phi+\frac{1}{2}\kappa\phi^2+\frac{1}{3!}\lambda\phi^3-g_s\rho_s,\\
&\nabla^2& V_0=m_v^2 V_0 -g_v\rho_B,\\
&\nabla^2& b_0=m_\rho^2 b_0-\frac{g_\rho}{2}\rho_3,
\end{eqnarray}
where $\rho_s$, $\rho_B$ and $\rho_3$ are defined next.
By replacing the meson fields by their mean values
\begin{eqnarray}
\phi&\rightarrow& \langle\phi\rangle=\phi_0,\\
V_\mu&\rightarrow& \langle V_\mu\rangle=V_0,\\
b_\mu&\rightarrow& \langle b_\mu\rangle=b_0,
\end{eqnarray}
the equations of motion read

\begin{eqnarray}
\phi_0&=&-\frac{\kappa}{2m_s^2}\phi_0^2-\frac{\lambda}{6m_s^2}\phi_0^3+\frac{g_s}{m_s^2}\rho_s,\\
V_0&=&\frac{g_v}{m_v^2}\rho_B,\\
b_0&=&\frac{g_\rho}{2m_\rho^2}\rho_3,
\end{eqnarray}
where $\rho_B=\rho_p+\rho_n$ is the baryonic density and $\rho_3=\rho_p-
\rho_n$, $\rho_p$ and $\rho_n$ are the
proton and neutron densities given by

\begin{equation}
\rho_j=2\int \frac{d^3p}{(2\pi)^3}(f_{j+}-f_{j-}), \quad j=p,n
\end{equation}
where $f_{j\pm}=1/(1+\exp{[(\epsilon_{j}\mp\nu_{j})/T]})$, $\epsilon_{j}=\sqrt{p^2+M^{*2}}$ and 
$\nu_{j}=\mu_{j}-g_vV_0-g_\rho \tau_3b_0$, where $\tau_3$ is the
appropriate isospin projector for the baryon charge states and $\mu_{j}$ are 
the nucleon chemical potentials. The scalar density $\rho_s$ is given by
\begin{equation}
\rho_s=2\sum_{j=p,n}\int \frac{d^3p}{(2\pi)^3}\frac{M^*}{\epsilon_j}(f_{j+}+f_{j-}).
\end{equation}

The thermodynamic quantities of interest are given in terms of the meson fields. They are the total energy density
\begin{equation}
\mathcal{E}_T=\mathcal{E}+\sum_{l=e,\nu}\mathcal{E}_l,
\end{equation}
with
\begin{eqnarray}
\mathcal{E}&=&\frac{1}{\pi^2}\sum_{j=p,n}\int dp~p^2\sqrt{p^2+M^{*2}}(f_{j+}+f_{j-})\\
&+&\frac{m_v^2}{2}V_0^2 +\frac{m_\rho^2}{2}b_0^2+\frac{m_s^2}{2}\phi_0^2\\
&+&\frac{\kappa}{6}\phi_0^3+\frac{\lambda}{24}\phi_0^4,
\end{eqnarray}
the total pressure is 
\begin{equation}
P_T=P+\sum_{l=e,\nu}P_l,
\end{equation}
with
\begin{eqnarray}
P&=&\frac{1}{3\pi^2}\sum_{j=p,n}\int dp~p^2\frac{p^4}{\sqrt{p^2+M^{*2}}}(f_{j+}+f_{j-})\\
&+&\frac{m_v^2}{2}V_0^2 +\frac{m_\rho^2}{2}b_0^2-\frac{m_s^2}{2}\phi_0^2\\
&-&\frac{\kappa}{6}\phi_0^3-\frac{\lambda}{24}\phi_0^4,
\end{eqnarray}
and the total entropy density
\begin{equation}
\mathcal{S}=\frac{1}{T}(\mathcal{E}_T+P_T-\sum_{j=p,n}\mu_j\rho_j-
\sum_{l=e,\nu}\mu_l\rho_l),
\end{equation}
where the electron and electron neutrino energy densities are
\begin{equation}
\mathcal{E}_l=\frac{g_l}{2 \pi^2}\int dp~p^2\sqrt{p^2+m_l^2}(f_{l+}+f_{l-}),
\end{equation}
and electron and electron neutrino pressure are
\begin{equation}
P_l=\frac{g_l}{6\pi^2}\int dp~\frac{p^4}{\sqrt{p^2+m_l^2}}(f_{l+}+f_{l-}).
\end{equation}
The electron density $\rho_e$ and electron neutrino density $\rho_\nu$ are given by
\begin{equation}
\rho_{l}=g_l\int \frac{d^3p}{(2\pi)^3}(f_{l+}-f_{l-}),\\
\end{equation}
where $g_e=2$, $g_\nu=1$, 
$f_{l\pm}=1/(1+\exp{[(\epsilon_l\mp\mu_l)/T]})$ with 
$\epsilon_l=\sqrt{p^2+m_l^2}$ and $\mu_e$ is the electron chemical potential,  
$\epsilon_\nu$ is the electron neutrino energy and $\mu_\nu$ is the electron 
neutrino chemical potential.
The condition of $\beta$ equilibrium in a system of protons, neutrons, 
electrons and trapped electron neutrinos is
\begin{equation}
\mu_p=\mu_n-\mu_e+\mu_\nu.
\end{equation}
We impose neutrality of charge as $\rho_p=\rho_e$ and fix the lepton fraction
\begin{equation}
Y_L=\frac{\rho_e+\rho_\nu}{\rho_B}.
\end{equation}

Notice that muons are not considered in the present calculation.

\section{Coexisting phases: neutral nuclear matter with neutrino trapping}

The formation of pasta phase has been studied lately with great interest
\cite{watanabe2005, horowitz2004}. Next we show the main steps for the 
calculation of the pasta phase with the coexistence phases method based on 
\cite{Barranco1980,Menezes1999}. For further details, please refer to 
\cite{pasta1,pasta2}.

For a given total density $\rho_B$ and lepton fraction $Y_L$ we
build pasta structures with different geometrical forms in a background
nucleon gas with $\beta$ stability and neutrino trapping. This is 
achieved by calculating from the Gibbs conditions
the density and the particle fractions of the pasta and of the background
gas so that in the whole we had to solve simultaneously the following
seven equations
\begin{eqnarray}
P^{I}(\nu_p^I,\nu_n^I,M^{*I})&=&P^{II}(\nu_p^{II},\nu_n^{II},M^{*II}),\\
\mu_n^I&=&\mu_n^{II},\\
\mu_e^I&=&\mu_e^{II},\\
\mu_\nu^I&=&\mu_\nu^{II},\\
m_s^2\phi_0^I+\frac{\kappa}{2}(\phi_0^I)^2+\frac{\lambda}{6}(\phi_0^I)^3&=&g_s\rho_s^I,\\
m_s^2\phi_0^{II}+\frac{\kappa}{2}(\phi_0^{II})^2+\frac{\lambda}{6}(\phi_0^{II})^3&=&g_s\rho_s^{II},\\
f(\rho_p^I-\rho_e^I)+(1-f)(\rho_p^{II}-\rho_e^{II})&=&0,
\end{eqnarray}
where $I$ and $II$ label each of the phases, $f$ is the volume fraction of
phase $I$
\begin{equation}
f=\frac{\rho_B-\rho^{II}}{\rho^{I}-\rho^{II}}.
\end{equation}
The total pressure is given by $P_T=P^I+P_e+P_\nu$. The total energy density
of the system is given by
\begin{equation}
\mathcal{E}=f\mathcal{E}^I+(1-f)\mathcal{E}^{II}+\mathcal{E}_e+\mathcal{E}_\nu+\mathcal{E}_{surf}+\mathcal{E}_{Coul},
\end{equation}
with $\mathcal{E}_{surf}=2\mathcal{E}_{Coul}$
\cite{Ravenhall1983,Maruyama2005}, and
\begin{equation}
\mathcal{E}_{Coul}=\frac{2\alpha}{4^{2/3}}(e^2\pi\Phi)^{1/3}[\sigma D(\rho_p^I-\rho_p^{II})]^{2/3},
\end{equation}
where $\alpha=f$ for droplets, rods and slabs and $\alpha=1-f$ for bubbles and
tubes, $\sigma$ is the surface energy
coefficient, $D$ is the dimension of the system.
For droplets, rods and slabs, $\Phi$ is given by
\begin{equation}
\Phi=\left\{
	\begin{array}{l}
	\left[\left(\frac{2-Df^{1-\frac{D}{2}}}{D-2}+f\right)\frac{1}{D+2}\right]
,\quad D=1,3,\\
	\frac{f-1-\ln(f)}{D+2},\quad  D=2,
	\end{array}
	\right.	
\end{equation}
and for bubbles the above expressions are valid with $f$ replaced by $1-f$.
The surface tension plays a significant role on the appearance of the pasta 
phase. In our treatment of the surface tension we essentially follow the 
prescription given in ~\cite{pasta1,pasta2}, but some comments on the
importance of the surface energy on the calculation of the pasta phase are
mandatory. It has been shown that the existence of 
the pasta phase as the lowest free energy matter and of its internal 
structures essentially depends on the value of the surface tension 
\cite{maruyama,watanabe2001,pasta1,pasta2,pasta_alpha}.
In the present paper the surface energy coefficient is parametrized in terms 
of the proton fraction according to the functional proposed in  
\cite{lattimer}, obtained by fitting Thomas-Fermi and Hartree-Fock numerical 
values with a Skyrme force. The same prescription was used in 
\cite{pasta1,pasta2}. However, a better recipe is to consider the surface 
energy coefficient in a consistent way, in terms of relativistic models. In 
\cite{pasta_alpha} the surface energy was parametrized according to
the Thomas-Fermi calculations for three parametrizations of the relativistic 
NLWM. The Gibbs prescription was used to obtain the $\sigma$ coefficient 
which is the appropriate surface tension coefficient to be used 
\cite{centel,mayers}. This improvement will be
added to our calculations in a forthcoming work. 

\section{Neutrino cross sections}

To calculate neutrino opacities and mean free paths we consider 
\cite{burrows1986} neutral current scattering reactions

\begin{eqnarray}
\nu_e+n\rightarrow\nu_e+n,\label{eq:scan}\\
\nu_e+p\rightarrow\nu_e+p,\label{eq:scap}
\end{eqnarray}
and charged current absorption reactions
\begin{eqnarray}
\nu_e+n\rightarrow e^-+p. \label{eq:absn}\\
\overline{\nu}_e+p\rightarrow e^++n.\label{eq:absp}
\end{eqnarray}

The cross sections for reactions (\ref{eq:scan}), (\ref{eq:scap}), 
(\ref{eq:absn}) and (\ref{eq:absp}) employed in this study follow 
\cite{burrows1986}:\\
Reaction (\ref{eq:scan}):
\begin{equation}\label{sigman}
\sigma_n=\left\{ \begin{array}{ll}	
\sigma_{ref}=\left(\frac{\sigma_0}{4}\right)\left(\frac{\epsilon_\nu}{m_ec^2}\right)^2, \qquad \mbox{$nND$, $\nu D$ or $\nu ND$},\\
\sigma_{ref}\left(\frac{\epsilon_\nu}{p_Fc}\right)\left(\frac{(1+4g_A^2)}{5}\right), \qquad \mbox{$nD$, $\nu ND$,~\cite{Iwamoto1981}},\\
\sigma_{ref}\left(\frac{1}{2}\right)\left(\frac{\pi^2(1+2g_A^2)}{8}\right)\times\\
\left(\frac{T}{\epsilon_\nu}\right)\left(\frac{T}{p_Fc}\right)\left(\frac{M^*c^2}{\epsilon_F}\right), \qquad\mbox{$nD$, $\nu D$,~\cite{goodwin1982,sawyer1979}}.
\end{array} \right.
\end{equation}
Reaction (\ref{eq:scap}):
\begin{equation}\label{sigmap}
\sigma_p=\left\{ \begin{array}{ll}	
\sigma_n, & \mbox{$pND$, $\nu D$ or $\nu ND$},\\
\sigma_n\left(\frac{Y_n}{Y_p}\right)^{1/3}, & \mbox{$pD$, $\nu ND$},\\
\sigma_n\left(\frac{Y_n}{Y_p}\right), &\mbox{$pD$, $\nu D$,~\cite{goodwin1982}}.
\end{array} \right.
\end{equation}
Reactions (\ref{eq:absn}) and (\ref{eq:absp}):
\begin{equation}\label{sigmaa}
\sigma_a=\left\{ \begin{array}{ll}	
\sigma_{ref}(1+3g_A^2), \qquad \mbox{$nND$, $\nu ND$},\\
\sigma_{ref}(1+3g_A^2)\left(\frac{2Y_p}{Y_n+Y_p}\right), \qquad \mbox{$nND$, $\nu D$ or $\nu ND$~\cite{bludman1978}},\\
\sigma_{ref}(1+3g_A^2)\left(\frac{1}{2}\right)\left(\frac{3\pi^2}{16}\right)\times\\
\qquad\left(\frac{T}{\epsilon_\nu}\right)^2\left(\frac{M^*c^2}{\epsilon_F}\right)\left(\frac{Y_e}{Y_n}\right)^{1/3},\qquad\mbox{$nD$, $\nu D$,~
\cite{Iwamoto1981}},\\
0, \qquad\mbox{$nD$, $Y_L<0.08$}.
\end{array} \right.
\end{equation}

In this expressions $p_F$ and $\epsilon_F$ mean the Fermi momentum and Fermi energy of
the degenerate neutron. $Y_e$, $Y_n$, $Y_p$, $Y_L$, are the electron, neutron, 
proton and lepton fractions.
ND denotes the non degenerate regime, while D denotes the validity in case of 
degenerate particles. 
$\sigma_0=1.76\times 10^{-44}~cm^2$ and $g_A=1.254$.	
Regions of intermediate degeneracy are also handled: the degenerate and
non degenerate sectors for both the baryons and the neutrinos of the cross 
sections detailed in equations (\ref{sigman}), (\ref{sigmap}) and 
(\ref{sigmaa}) are joined by a simple interpolation algorithm as done 
in~\cite{keil1995}.
The total neutrino mean free path in dense matter is given by
\begin{equation}
\lambda_\nu=\frac{1}{\rho_n\sigma_n+\rho_p\sigma_p+\rho_B\sigma_a}.
\end{equation}
Rosseland neutrino mean free paths are related with diffusion coefficients $D_j$~\cite{pons1999} by
\begin{equation}
\lambda_\nu^k=\frac{D_k}{\int_0^{\infty}d\epsilon_\nu~\epsilon_\nu^k f_\nu
(1-f_\nu)},
\end{equation}
where
\begin{equation}\label{eq:diff}
D_k=\int_0^{\infty}d\epsilon_\nu~\epsilon_\nu^k \lambda_\nu f_\nu(1-f_\nu)
. \quad k=2,3,4
\end{equation}
All contributions from neutrino opacities are related with the diffusion 
coefficients and can to be used as input to the solution of the transport 
equations in the equilibrium diffusion approximation to simulate the 
Kelvin-Helmholtz phase of the protoneutron stars \cite{pons1998}.

\section{Results and Conclusions}

\begin{figure}
\caption{\label{fig:f504} free
energy per particle with the NL3 parametrization
obtained for $T = 5 MeV$ and $Y_L=0.4$.}
\includegraphics[scale=0.32]{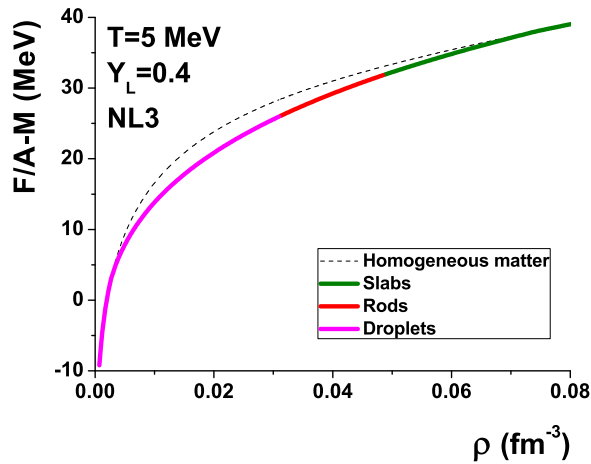}
\end{figure}

Before we tackle the problem of the consequences of the pasta phase on the
diffusion coefficients, we display a characteristic  figure of the
free energy for the homogeneous and pastalike matter obtained for $T=5$ MeV and
$Y_L=0.4$ in Fig. \ref{fig:f504}. A similar figure is presented in Fig. 9 of
\cite{pasta_alpha}, but obtained with a relativistic surface energy. One can
see that the pasta phase ends when the free energy density reaches the curve
for the homogeneous matter. Actually, at this temperature, the pasta phase
interpolates between two regions of homogeneous matter, which is the
preferential ground state at extremely low densities, as seen in 
Fig.\ref{fig:f504}.
Moreover, the size of the pasta phase decreases
with the increase of the temperature and eventually, it no longer exists.
It is also worth mentioning that neutrino free matter in $\beta$-equilibrium
presents a pasta phase smaller than matter with trapped neutrinos 
\cite{pasta2,pasta_alpha} as a consequence of the fact that the latter presents
a larger fraction of protons. According to studies on binodals and spinodals 
underlying the conditions for phase coexistence and phase transitions 
\cite{camille08,cphelena,pasta1}, non-homogeneous matter with trapped neutrinos 
is expected to be found until temperatures around $T=12$ MeV, depending on 
the model considered.

We next show the diffusion coefficients $D_2,D_3$ and $D_4$ as function of the 
baryon density for different temperatures obtained for both homogeneous 
matter and the pasta phase. According to \cite{pasta1,pasta2} the densities 
where matter becomes homogeneous depend on the proton fraction and on the 
temperatures involved, but it is always smaller than 0.1 fm$^{-3}$ for the NL3 
parametrization and for the $\sigma$ values we consider in the present work.

In obtaining the diffusion coefficients, the EoS was calculated as a grid where
temperature ranges are in between 0 and 50 MeV and densities vary from 0.005 to
0.5 fm$^{-3}$. In our codes we have implemented a prescription given in 
\cite{aparicio} to evaluate the Fermi integrals so that the same codes run 
from zero ($10^{-9}$) to high temperatures. 
We have calculated the diffusion coefficientes only for baryonic 
densities above $0.005~fm^{-3}$ because the integrals of type (\ref{eq:diff}) 
are very difficult to converge at lower densities.
We show results for lepton fractions
equal to 0.2 and 0.4 because those are typical values necessary in the numerical
simulation of protoneutron star evolution.

In all figures the diffusion coefficients obtained with homogeneous matter
join the curves obtained with the pasta phase at densities higher than the 
ones shown. For $D_2$ calculated at T=5 MeV and $Y_L=0.4$, for instance, they
cross each other at $\rho=0.12$ fm$^{-3}$. Our codes interrupt the calculation
once homogeneous matter becomes the ground state configuration, as depicted
in Fig. \ref{fig:f504}. This means that there will always be a gap in the 
diffusion coefficients when the transport equations are calculated with the
inclusion of the pasta phase. The same behavior is found at the pressure 
values for homogeneous and pasta phases at the transition density.

\begin{widetext}
\begin{figure}
\caption{\label{fig:D2402} Diffusion coefficient $D_2$ as function of baryon density for different temperature and proton fraction values for homogeneous matter and pasta phase.}
\includegraphics[scale=0.32]{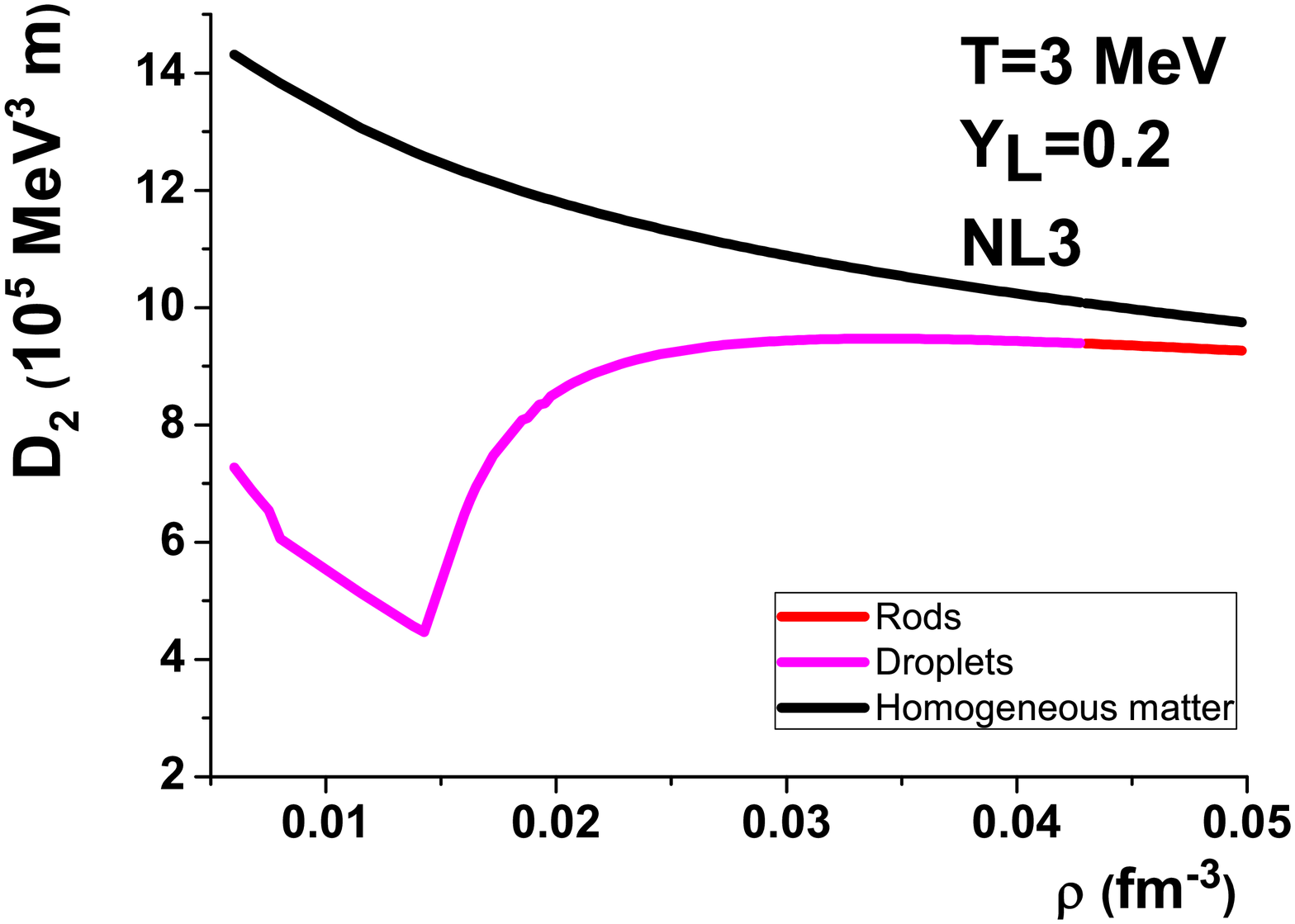}\includegraphics[scale=0.32]{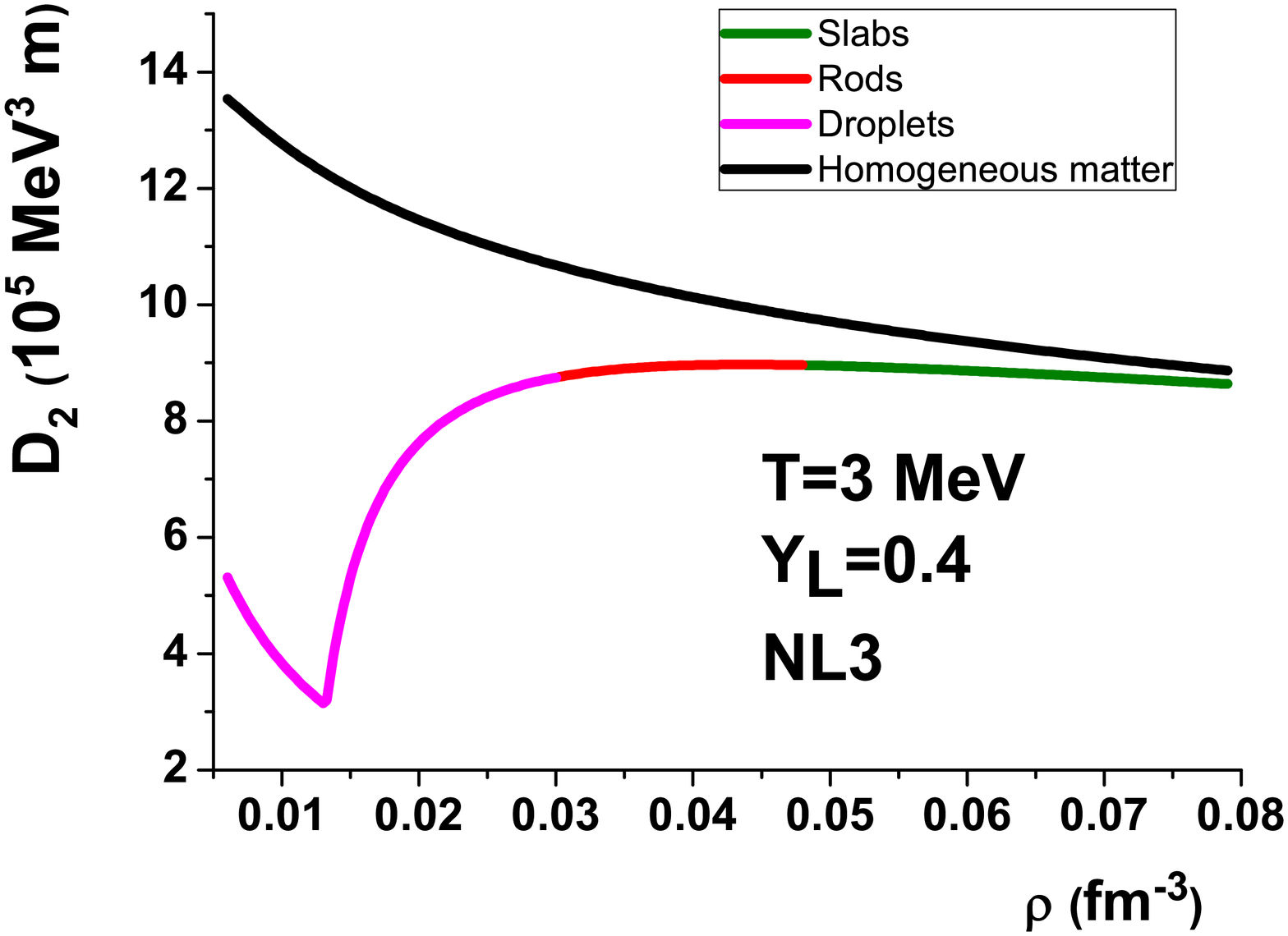}\\
\includegraphics[scale=0.32]{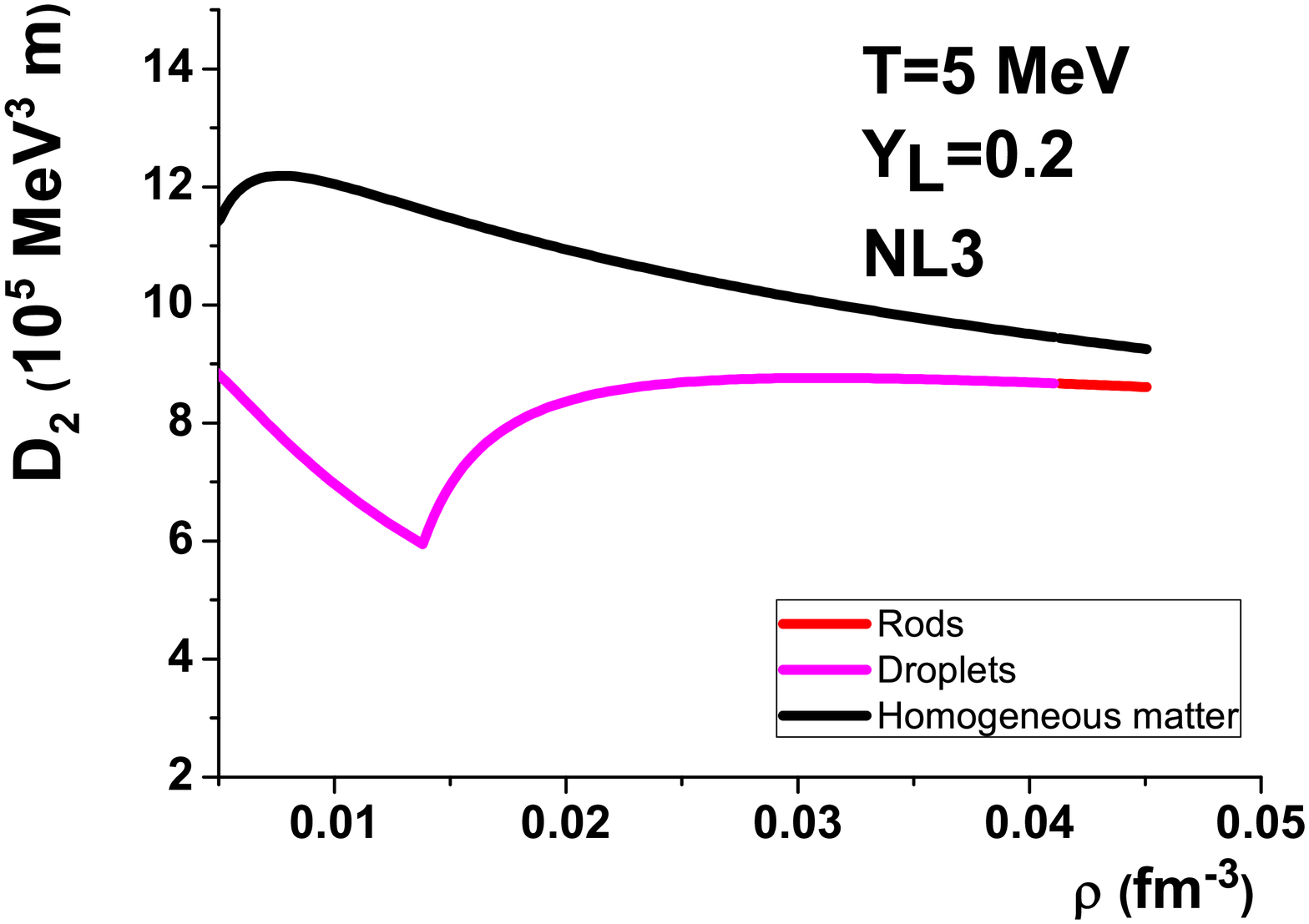}\includegraphics[scale=0.32]{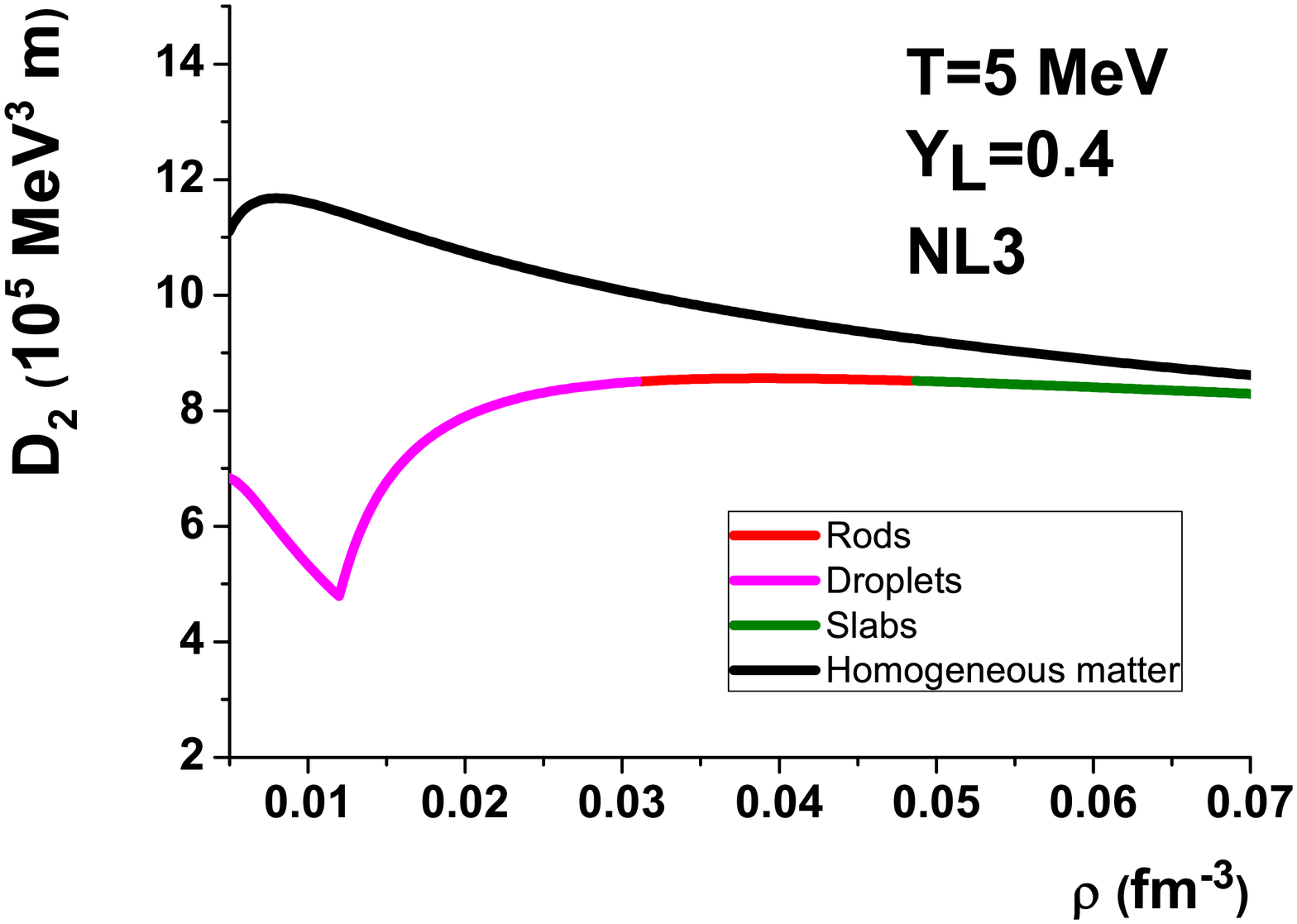}\\
\end{figure}
\end{widetext}

\begin{widetext}
\begin{figure}
\caption{\label{fig:D2402b} Diffusion coefficient $D_3$ as function of baryon density for different temperature and proton fraction values for homogeneous matter and pasta phase.}
\includegraphics[scale=0.32]{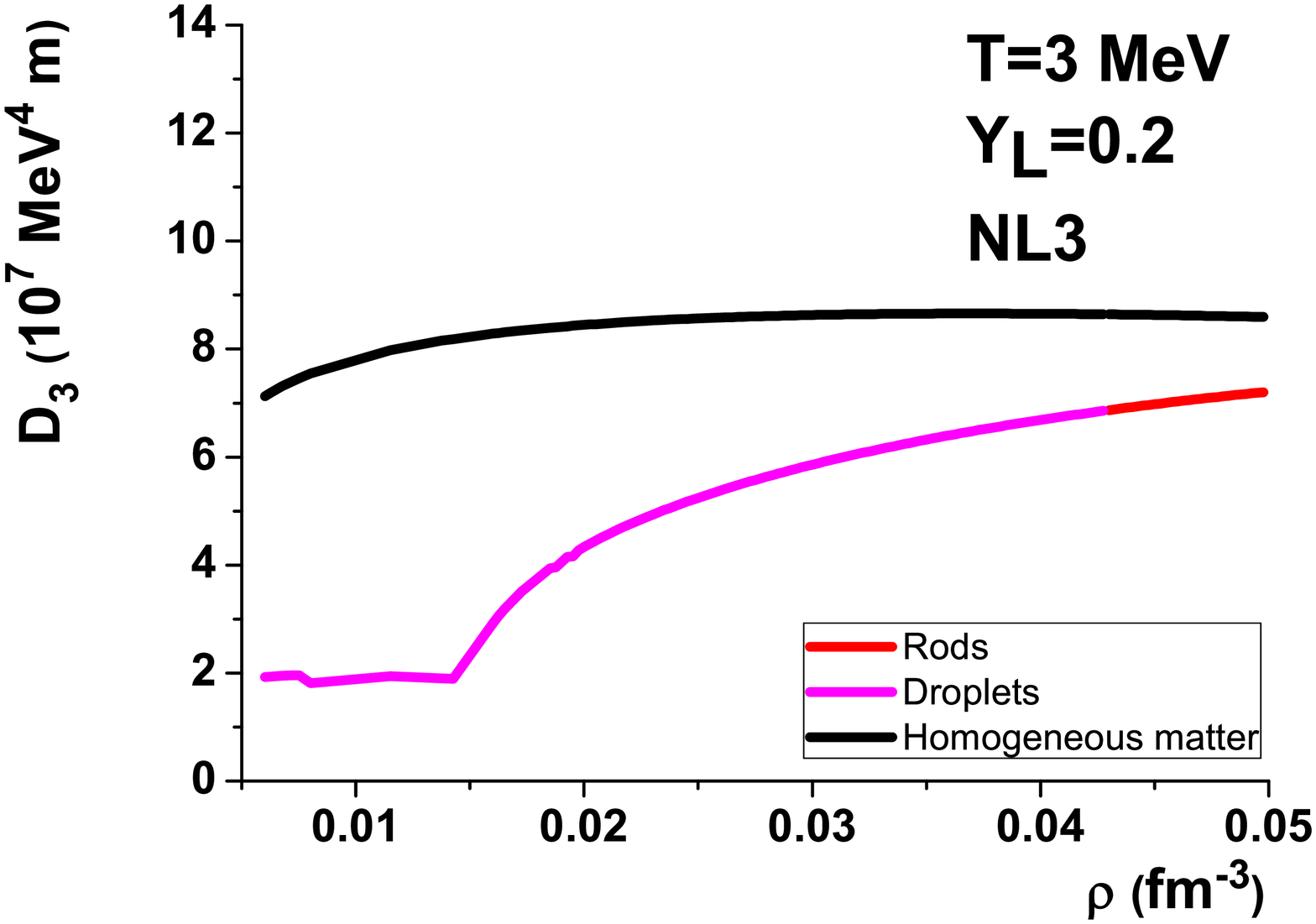}\includegraphics[scale=0.32]{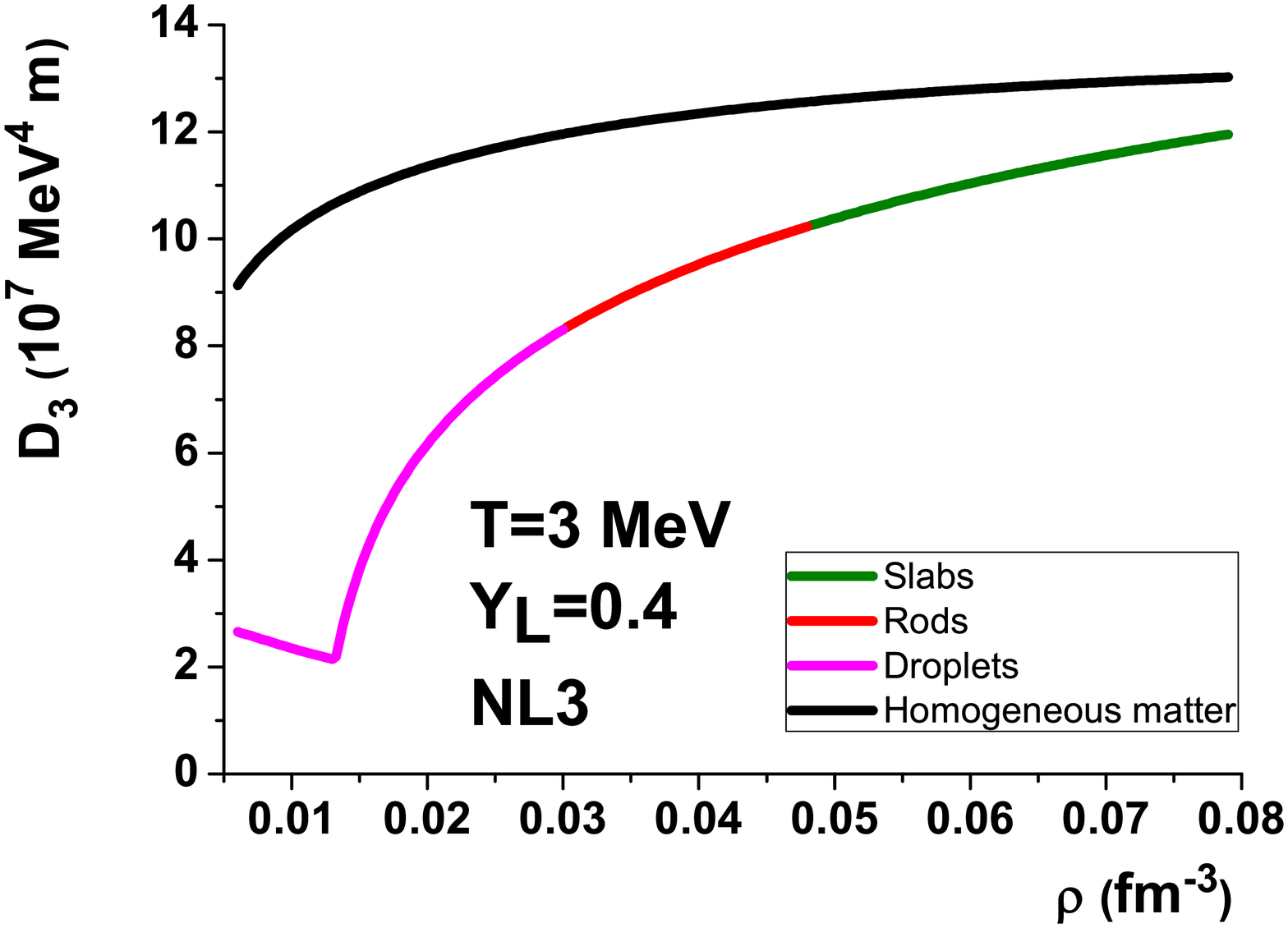}\\
\includegraphics[scale=0.32]{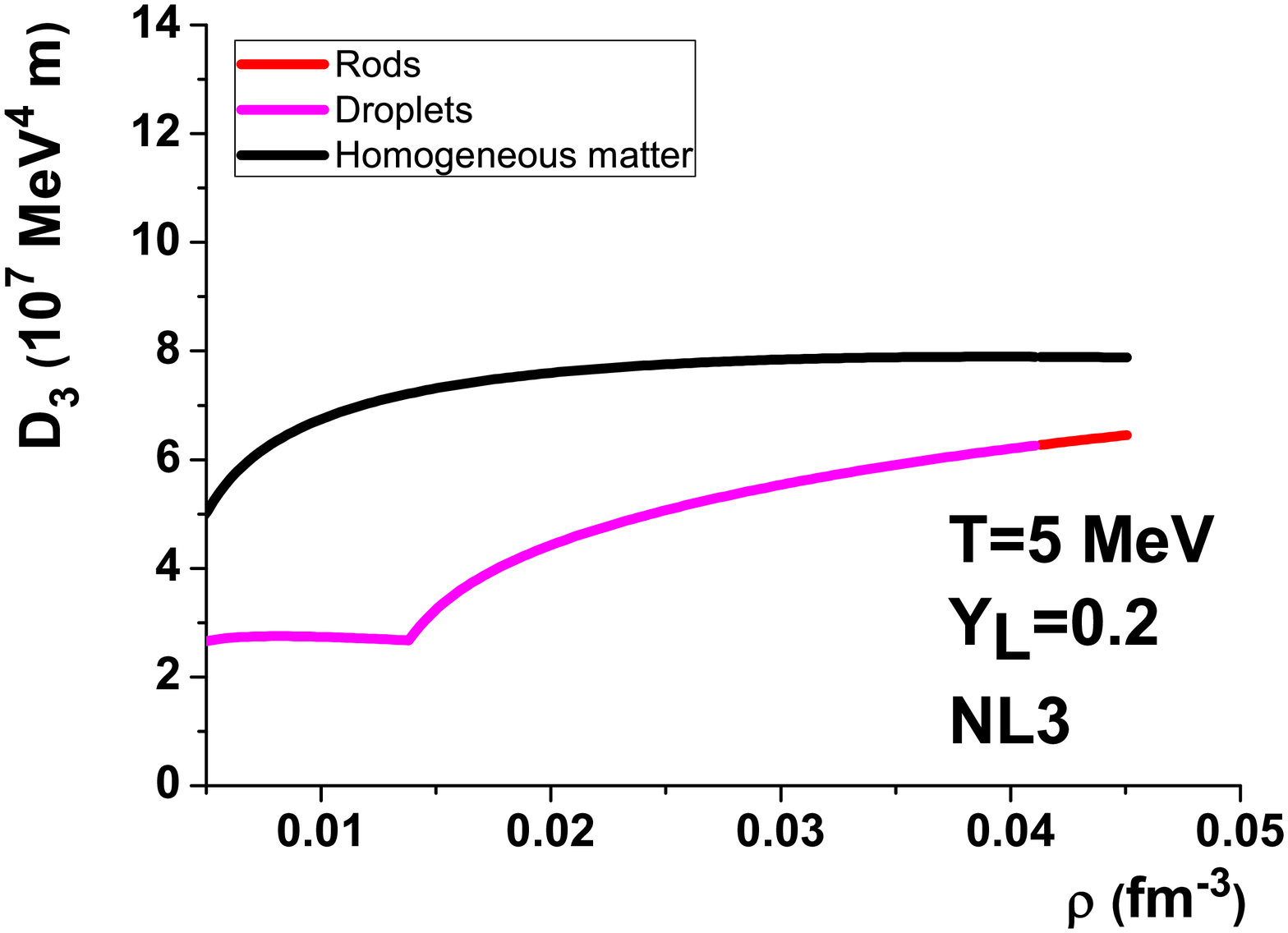}\includegraphics[scale=0.32]{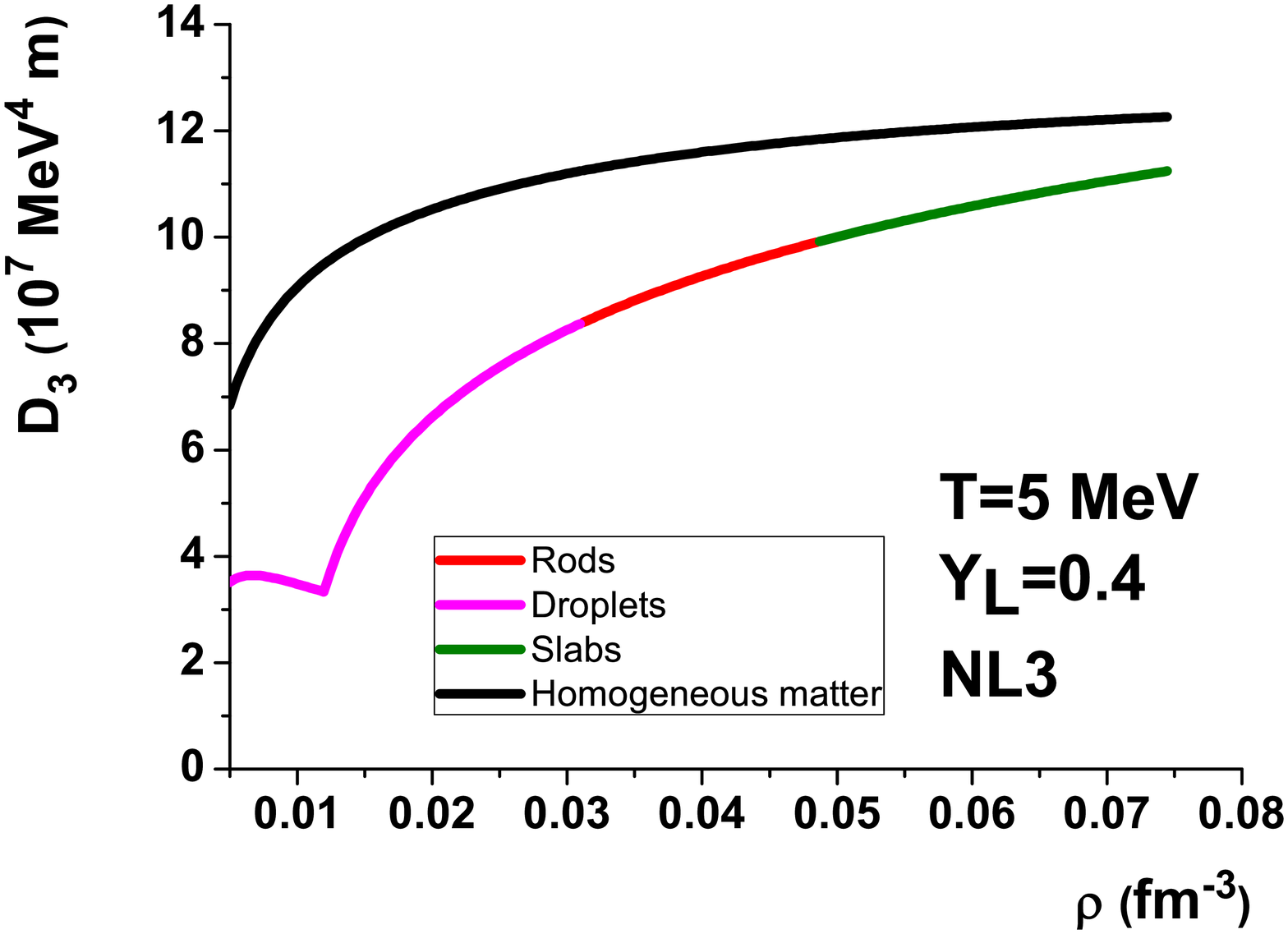}\\
\end{figure}
\end{widetext}

\begin{widetext}
\begin{figure}
\caption{\label{fig:D2402c} Diffusion coefficient $D_4$ as function of baryon density for different temperature and proton fraction values for homogeneous matter and pasta phase.}
\includegraphics[scale=0.32]{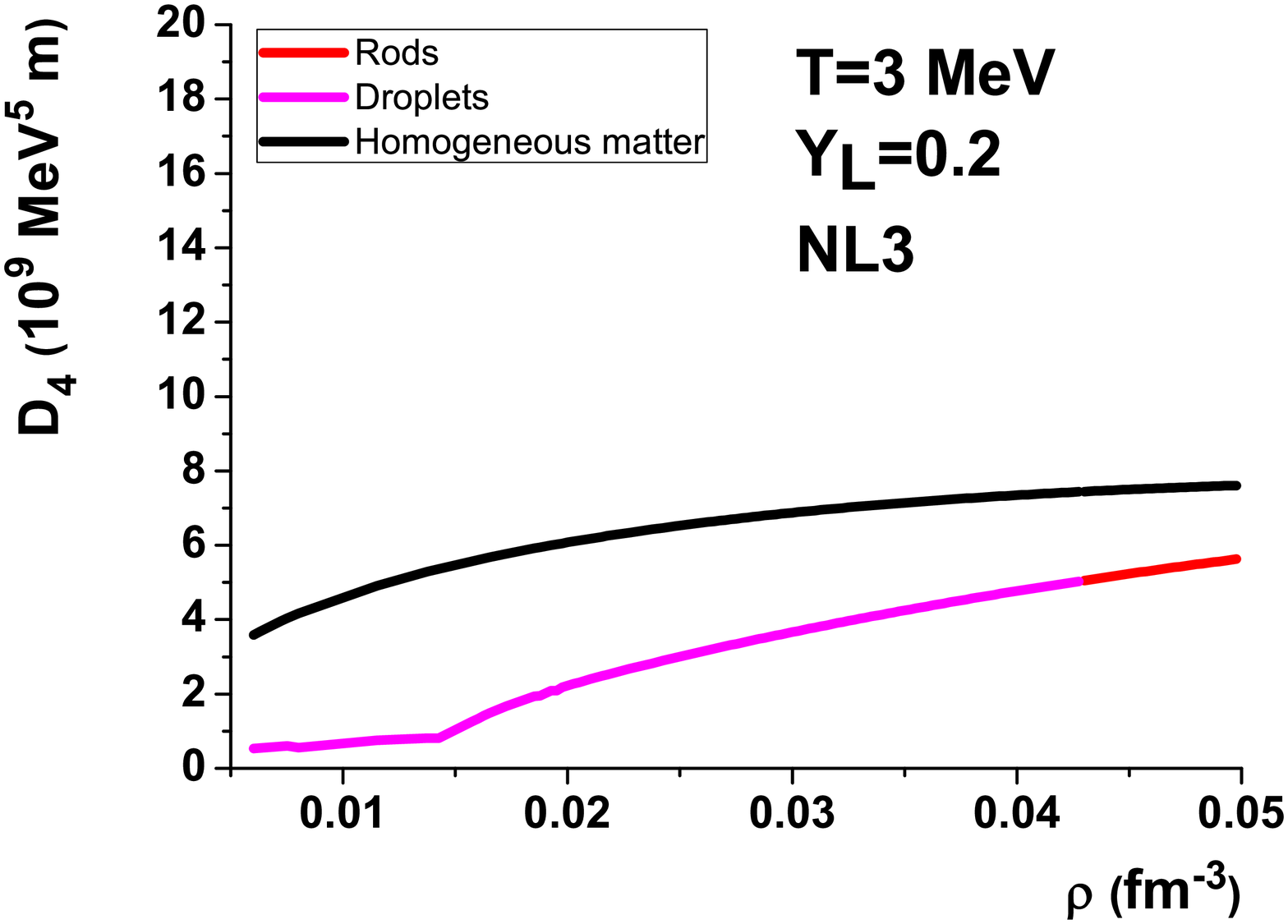}\includegraphics[scale=0.32]{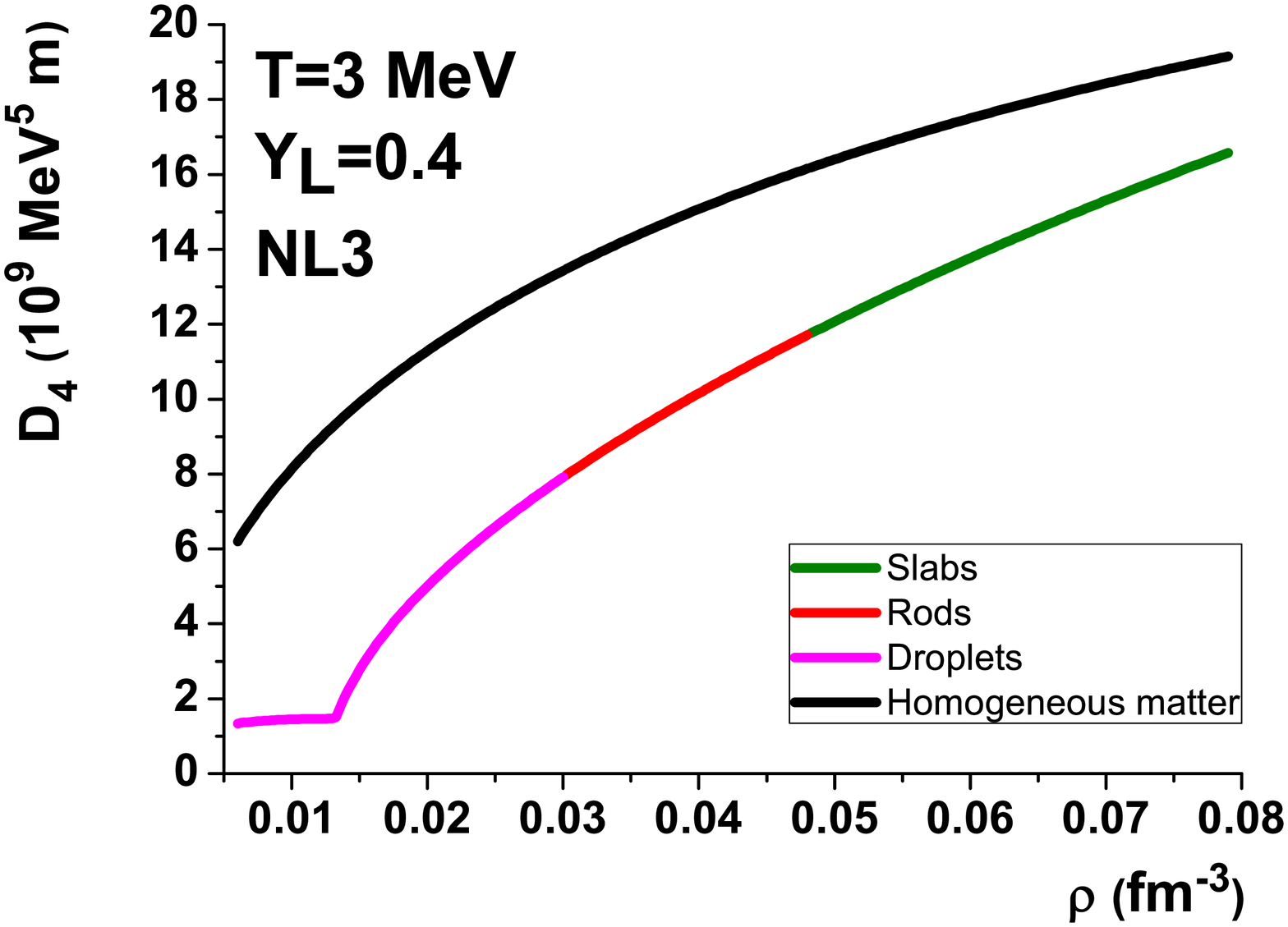}\\
\includegraphics[scale=0.32]{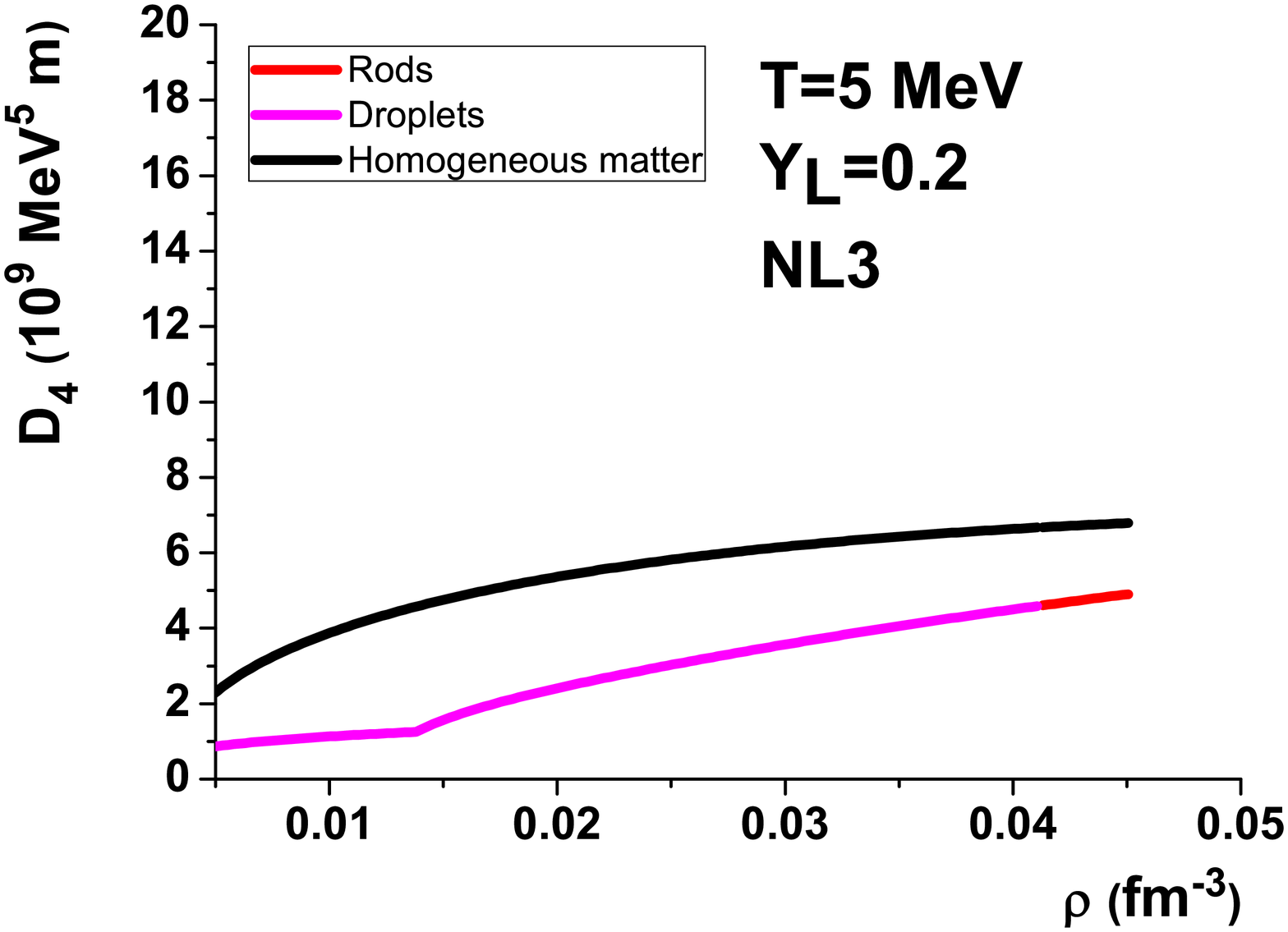}\includegraphics[scale=0.32]{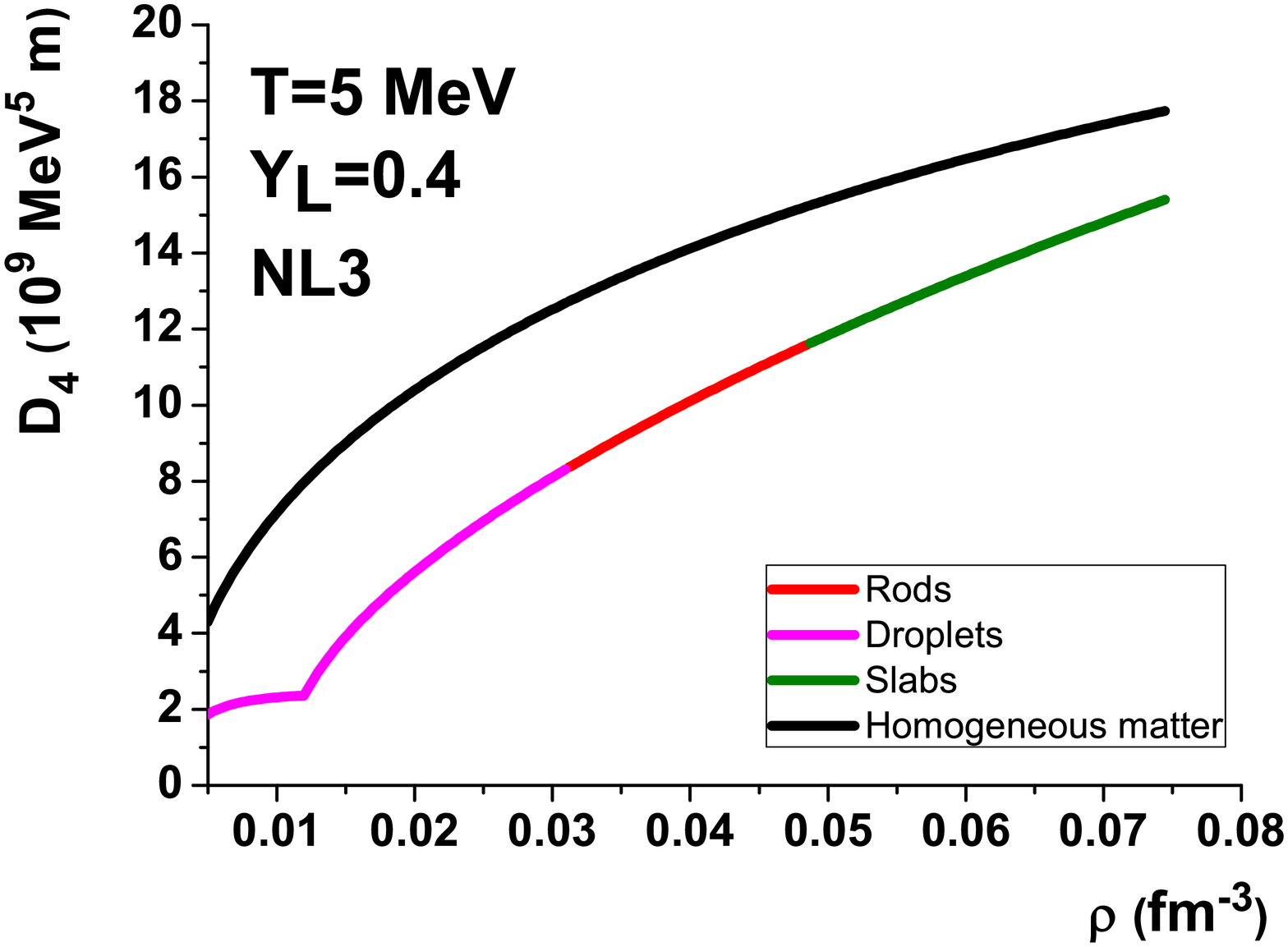}\\
\end{figure}
\end{widetext}

From figures \ref{fig:D2402}, \ref{fig:D2402b} and \ref{fig:D2402c} 
we can see that only three structures are found inside the pasta phase for the 
present model: droplets, rods and slabs as far as $Y_L=0.4$. For $Y_L=0.2$
only the first two structures remain. While the diffusion coefficients obtained
with homogeneous matter is always smooth and continuous, a common trend of 
all the diffusion coefficients obtained with the pasta phase is a kink at very 
low densities in between 0.01 and 0.015 fm$^{-3}$. The interpolation 
procedure we use depends on the quantities $\eta_i=(\mu_i-M^*)/T, i=p,n$.
Whenever either $\eta_p$ or $\eta_n$ inverts its sign, these kinks 
appear, i.e., they are the result of the effective nucleon mass being greater 
than the corresponding chemical potential.
Moreover, the pasta phase diffusion coefficients  are always
lower than the corresponding coefficients obtained with homogeneous matter.

Our results for the diffusion coefficients $D_2$ and $D_4$ are one order of 
magnitude larger than the results obtained in~\cite{pons1998}. This difference 
can be explained because in the present paper all diffusion coefficients are 
calculated at very low baryonic densities. For larger densities the results 
coincide. 

In summary we point out that in the present paper we have investigated 
the influence of the pasta phase 
on the neutrino opacity by calculating the diffusion coefficients. The 
homogeneous EoS was obtained with the NL3 parametrization of the NLWM in a mean
field approximation. The pasta phase was obtained with the coexistence 
phases method (CP). 

Recent calculations for the pasta phase within the 
Thomas-Fermi approximation at finite temperatures \cite{pasta_warm} show that
the internal pasta structure is much richer as compared with the CP method we
have employed in the present work. Hence, the dependence on the structure of 
the pasta phase is also of interest and this calculation is planned for 
different parametrizations of the NLWM. More sophisticated matter, which 
includes the $\alpha$-particles should also be considered \cite{pasta_alpha}.

We have checked that the neutrino interactions in warm and low baryonic 
densities with pasta formation show significant differences when compared 
with homogeneous matter. Next the temporal evolution of the PNS will be
 calculated and, in face of the present results, we expect that the 
cooling and deleptonization eras will be influenced by the presence of the 
pasta phase at low densities.

An obvious improvement is the inclusion of hyperons in the EoS. However,
the pasta phase can still be computed just with protons, neutrons and light
clusters because hyperons are expected to appear only at densities where the 
pasta phase is no longer present.

\section*{ACKNOWLEDGMENTS}
This work was partially supported by the Brazilian sponsoring bodies 
CNPq  and CAPES (M.D. Alloy scholarship).

\newpage

\newpage

\newpage

\end{document}